\documentclass[%
 reprint,
superscriptaddress,
 amsmath,amssymb,
 aps,
prb,
]{revtex4-2}
\usepackage{float} 
\usepackage{booktabs}
\usepackage{mathrsfs}
\usepackage{amssymb}
\usepackage{amsmath,empheq}
\usepackage{graphicx}
\usepackage{dcolumn}
\usepackage{bm}
\usepackage{hyperref}
\usepackage{orcidlink}
\hypersetup
{
hypertex=true,
colorlinks=true,
linkcolor=blue,
filecolor=blue,
urlcolor=blue,
citecolor=blue,
breaklinks=true
}

\begin{document}

\preprint{AIP/123-QED}

\title{A low-temperature entropy source for on-chip true random number generation: universal robustness beyond device quality}

\author{Y. Q. Chai}
\altaffiliation{These authors contributed equally to this work}
\affiliation{HergD collaboration and Information Quantum Technology Laboratory, School of Information Science and Technology, Southwest Jiaotong University, Chengdu 610031, China}
\author{M. Y. Wang}
\altaffiliation{These authors contributed equally to this work}
\affiliation{HergD collaboration and Information Quantum Technology Laboratory, School of Information Science and Technology, Southwest Jiaotong University, Chengdu 610031, China} 
\author{X. N. Feng}
\affiliation{HergD collaboration and Information Quantum Technology Laboratory, School of Information Science and Technology, Southwest Jiaotong University, Chengdu 610031, China} 
\author{L. F. Wei}
\email{lfwei@swjtu.edu.cn}
\affiliation{HergD collaboration and Information Quantum Technology Laboratory, School of Information Science and Technology, Southwest Jiaotong University, Chengdu 610031, China}

\date{\today}
%
\begin{abstract}
True random number generation is a critical capability for fault-tolerant quantum computing at millikelvin temperatures. Yet existing Josephson-junction-based TRNGs all rest on a widely accepted but untested assumption: that reliable entropy extraction requires precisely controlled device parameters. Here we show that this assumption does not always hold. We demonstrate a counterintuitive finding: a single current-biased Josephson junction, regardless of its parameter quality, can serve as a cryptographic-grade true random number generator. To establish the universality of this conclusion, we deliberately selected the most extremely deviated devices from fabrication, with critical currents three orders of magnitude away from theoretical predictions and $I_cR$ products an order of magnitude above conventional values, as the ultimate stress test. Even under these extreme conditions, the raw Shannon entropy reaches 0.9981~bit (99.8\% of the theoretical maximum), with a min-entropy of 0.9271~bit. Using a square-wave pulsed-bias scheme, we tune the switching probability to $P\approx0.5$. After SHA-256 post-processing, the bitstreams pass all 15 NIST SP 800-22 tests under a conservative $m=3$ criterion that is more demanding than the standard recommendation, and this certification holds across the entire 100-700~mK operating window of a dilution refrigerator.
\end{abstract}

\maketitle
\section{Introduction}
The advent of fault-tolerant quantum computing creates an unprecedented demand for true random numbers on the dilution-refrigerator stage. In leading quantum error correction (QEC) architectures, such as surface codes that have recently surpassed the breakeven threshold on superconducting processors~\cite{Acharya2025}, randomness is a continuous, real-time operational necessity~\cite{menezes1996handbook}. Randomized compiling consumes fresh random bits per gate cycle to transform coherent errors into benign stochastic noise~\cite{PhysRevA.94.052325,PhysRevX.11.041039}; scaled to hundreds of qubits, the consumption grows with circuit volume, with recent implementations requiring per-cycle, per-qubit Pauli sampling directly on the control FPGA~\cite{hardware2024randomized}. Dynamic decoding and syndrome extraction demand microsecond-level random seed updates against side-channel attacks~\cite{Erata_Xu_Piskac_Szefer_2024}. These bits must be generated on demand, with minimal latency and sufficient unpredictability for cryptographic applications, as buffering or remote generation can introduce delays and additional security vulnerabilities. This urgency is underscored by recent Nature demonstrations of certifiably perfect randomness via superconducting qubits~\cite{Kulikov2026} and the strategic importance of random numbers for quantum-secure communications~\cite{Acin2016}.

Cryogenic physics imposes a severe bottleneck: conventional CMOS entropy sources fail below 4~K, with power dissipation prohibitive for mK thermal budgets~\cite{fu2021overview}. Superconducting digital control electronics, while integrable with qubits via flip-chip bonding at millikelvin temperatures~\cite{Jordan2026b}, still lack a dedicated on-chip entropy source, creating an urgent unmet need for a cryogenic-native, nanosecond-responsive, monolithically integrable true random number generator (TRNG).

Over the past decade, entropy sources from thermal noise~\cite{jun1999intel} to quantum measurements~\cite{ma2016quantum} have been explored. Josephson junctions (JJs) are natural candidates for cryogenic entropy sources~\cite{PhysRevResearch.6.013236}. Yet existing JJ-based TRNGs converge on two architectures incompatible with real-time QEC. The single-flux-quantum (SFQ)-logic approach converts stochastic switching into digital pulses via inductor/resistor/multi-junction networks~\cite{zhou200150}, consuming substantial area, offering narrow margins, and proving highly sensitive to fabrication variances, as recent integration efforts show~\cite{Jordan2026b}. RF-driven chaotic approaches rely on external microwave sources, careful impedance matching, and extensive digitization, introducing latency and dependence on room-temperature electronics that can hinder on-chip integration~\cite{oikawa2024chaotic}. Neither delivers the just-in-time, low-latency, manufacturable entropy source fault-tolerant QEC demands. State-transition nanodevices remain thermally incompatible with cryogenic operation and suffer fatigue and fabrication sensitivity~\cite{gong2019true}.

Here we demonstrate a radically simpler paradigm: a single standalone CBJJ, free of SFQ periphery or RF excitation, constitutes a TRNG that inherently meets real-time QEC demands. By directly digitizing the junction's stochastic phase-escape, governed by thermal activation or macroscopic quantum tunneling~\cite{clarke1988quantum,devoret1985measurements}, we generate bits at the junction's natural switching rate, limited only by the Josephson plasma frequency, eliminating all intermediate buffering and pre-processing latency for nanosecond entropy response. Remarkably, this stochastic digitization is robust to macroscopic structural non-idealities, shadowing effects from Dolan evaporation, and parasitic variations that would disable SFQ designs. Junctions spanning substantial variations in critical current, parasitic capacitance, and resistance exhibit comparable entropy-generation performance, without the need for device screening, trimming, or post-selection. This tolerance to fabrication variability could enable higher-yield production and relax parameter-control requirements.

Our minimalist square-wave current-pulse scheme modulates switching probability via amplitude control, with low-frequency pulsed bias suppressing thermal accumulation for long-term stationarity. Systematic characterization from 100~mK to 700~mK confirms robust entropy across broad operating margins. Raw bitstreams, refined by lightweight SHA-256 post-processing, pass all NIST SP~800-22 tests with ample margins. The generator occupies a single-junction footprint, dissipates virtually zero static power, introduces no clock-domain overhead, and is a plug-and-play, always-on entropy source that natively interleaves with qubit control lines and readout resonators inside the dilution refrigerator. This stands in stark contrast to entanglement-assisted or multi-component optical quantum randomness schemes~\cite{Kulikov2026,Kavuri2025}, which achieve certified randomness but not qubit-level on-chip integration. Our work delivers the missing piece for fault-tolerant superconducting processors: a manufacturable, ultra-low-latency, cryogenic-compatible TRNG that feeds the insatiable randomness appetite of real-time QEC, finally decoupling quantum computing from the bandwidth, latency, and security limitations of room-temperature entropy banks.
\section{Result}
\subsection{Random switching in current-biased Josephson junctions}
A foundational assumption in superconducting electronics holds that reliable device functionality requires near-ideal device parameters. This assumption has guided decades of process optimization. For most applications, it is justified. For entropy generation, we show that it does not always hold. To test this premise, we deliberately selected from the Dolan double-angle evaporation process the devices with the most pronounced fabrication-induced deviations; the specimens that would normally be discarded in any SFQ circuit fabrication~\cite{osman2021simplified}. The rationale was straightforward: if a demonstrably ``bad'' junction yields reliable entropy, the argument for robustness becomes difficult to refute.

The switching mechanism itself is the junction's stochastic phase escape. Under a bias current $I_b<I_c$, the Josephson phase behaves like a particle in a tilted washboard potential; thermal fluctuations or macroscopic quantum tunneling can drive it over the barrier, switching the junction from the zero-voltage state to a finite-voltage resistive state~\cite{devoret1985measurements,clarke1988quantum}. For a bias pulse of duration $t_p$, the switching probability is $P_{sw}(I_b)=1-\exp[-\Gamma_{total}(I_b)t_p]\label{eq:P}$ (see Methods for the RCSJ model and explicit rate expressions)~\cite{mccumber1968effect,stewart1968current}. This behaviour is generic: every Josephson junction, whatever its critical current, capacitance or parasitic structure, displays the same phase-escape dynamics. A tenfold variation in $I_c$ does not suppress the escape process; it merely shifts the bias current needed to reach a given escape rate.

\begin{figure}[htbp]
\includegraphics[width=1\linewidth]{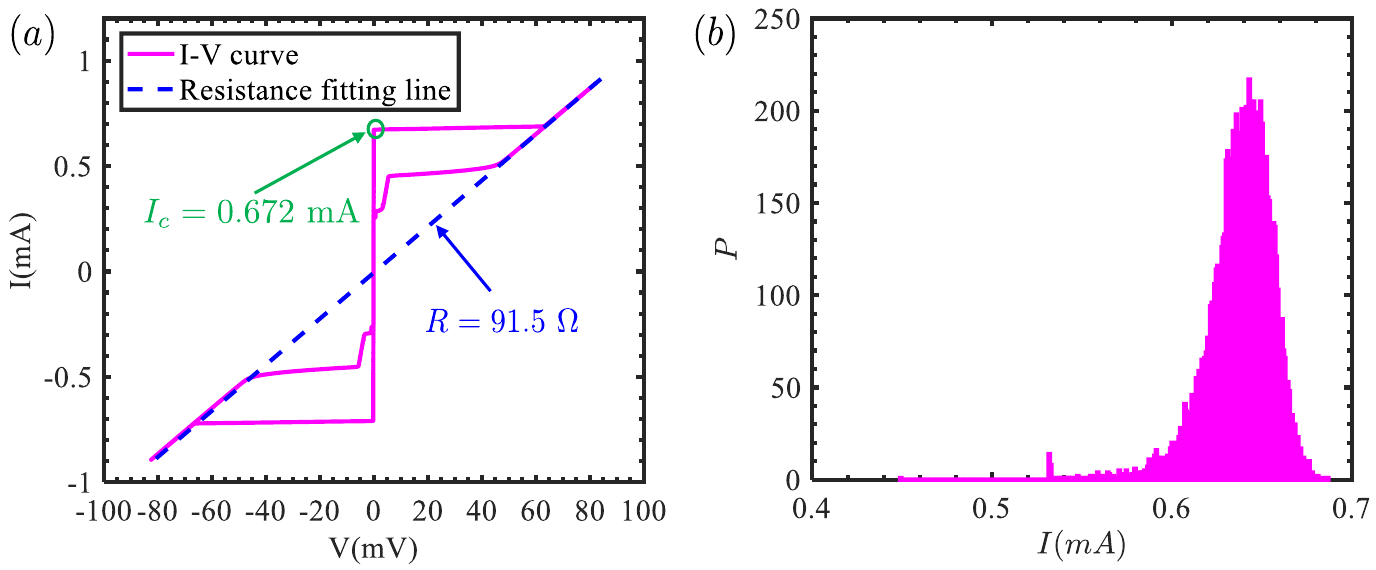}
\caption{\label{fig:5}
Electrical characteristics and stochastic switching behavior of the Josephson junction. (a) I–V characteristic of the JJ. The device exhibits significant underdamped hysteresis, with normal-state resistance and critical current of $91.5~\Omega$ and $0.672$~mA, respectively. (b) Switching current distribution measured at a constant current sweep rate.}
\end{figure}
We tested this universality directly with our non-ideal devices. Fig.~\ref{fig:5}(a) shows the measured current-voltage characteristic. The normal-state resistance $R=91.5~\Omega$ and critical current $I_c=0.672$~mA differ from the Ambegaokar-Baratoff prediction $\tilde{I}_c\approx3.09~\mu$A by three orders of magnitude~\cite{ambegaokar1963tunneling}; the $I_cR$ product of 61.25~mV exceeds typical Al/AlOx/Al values by an order of magnitude. These deviations, probably from shadowing-induced parasitic paths during evaporation, would immediately rule out the device for any SFQ circuit. Nevertheless, in 10,000 repeated switching measurements (Fig.~\ref{fig:5}(b)), the switching-current distribution showed a stable, broad width and was indistinguishable from that expected for an ideal junction~\cite{barone1982physics,chai2025measuring}. We observed no fatigue, no hysteresis, and no drift. For entropy generation, the ``bad'' junction behaves exactly like a ``good'' one.

\begin{figure}[htbp]
\includegraphics[width=0.9\linewidth]{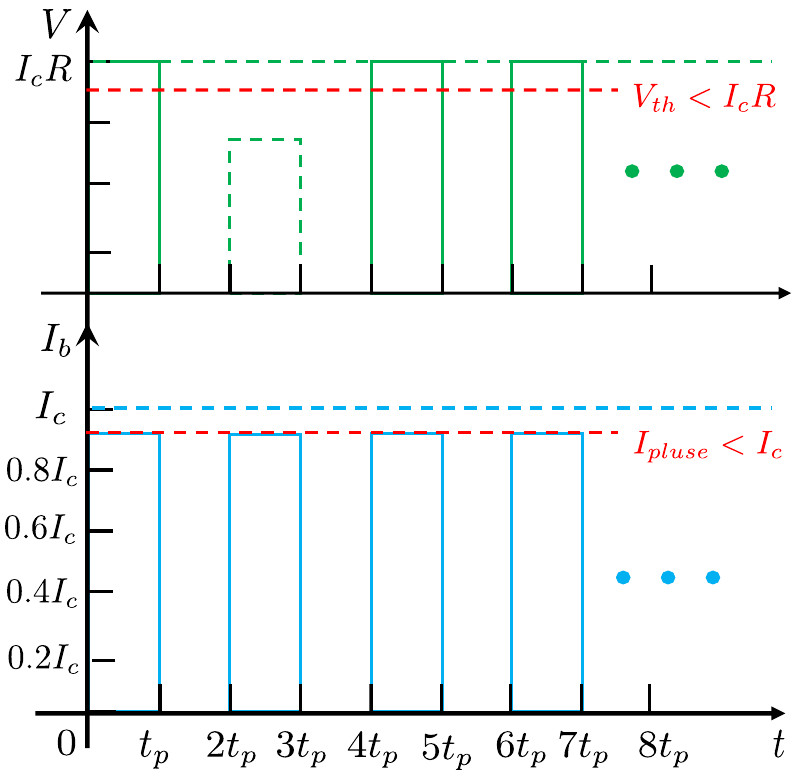}
\caption{\label{fig:2}
Schematic of the stochastic voltage switching of a CBJJ under a current pulse. The pulse duration and duty cycle are $t_p$ and 0.5, respectively. The pulse amplitude satisfies $I_{pulse}<I_c$, ensuring that voltage switching occurs only through noise-induced escape. When the CBJJ switches to the voltage state, the voltage across the junction is approximately $I_cR$. The voltage threshold $V_{th}$ for identifying a switching event is therefore set slightly below $I_cR$ to minimize false identification. The green solid and dashed curves represent the voltage responses of the JJ with and without switching, respectively, during a single current pulse.}
\end{figure}
\begin{figure}[htbp]
\includegraphics[width=1\linewidth]{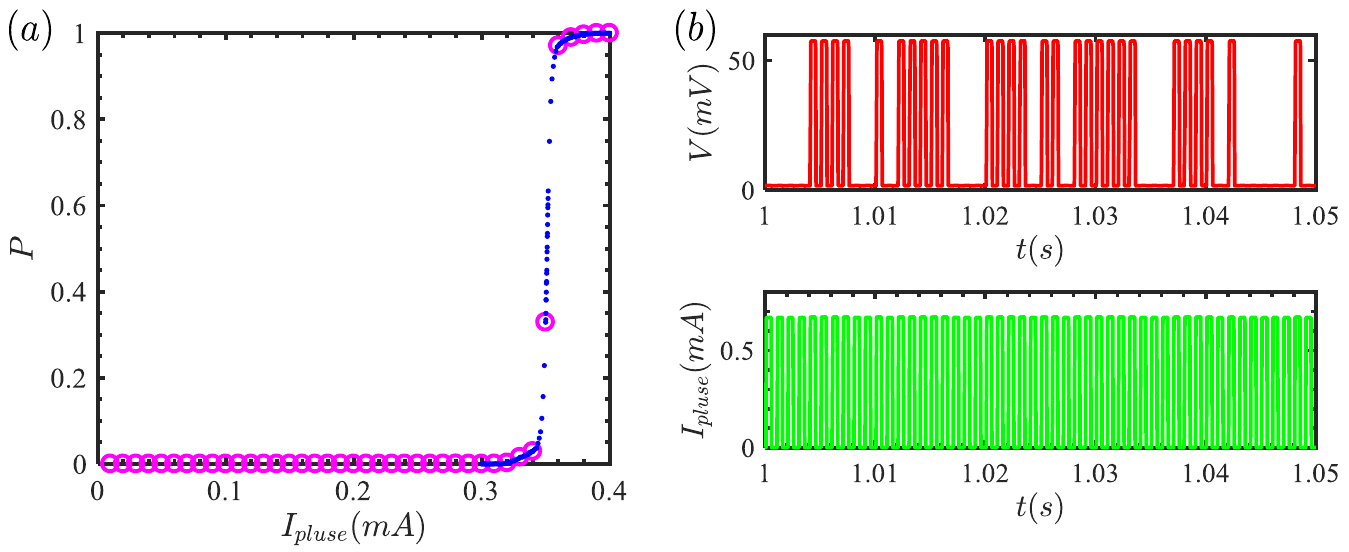}
\caption{\label{fig:7}
Dependence of JJ switching probability on pulse amplitude. (a) Variation of JJ switching probability with pulse amplitude under 100~mK and 1~kHz conditions. First, a larger step size is used to determine the switching region, then a fine scan is performed at a step length of $0.1~\mu$A. At $I_{ pulse}=0.3512$~mA, $P=0.5033$ is obtained. (b) Time-domain signals of bias current and JJ output voltage at optimal pulse amplitude, showing randomly generated ``0" and ``1" states.}
\end{figure}

We converted the stochastic switching into a binary stream using square-wave current pulses (Fig.~\ref{fig:2}) of duration $t_p=0.5$~ms and 50\% duty cycle. A threshold $V_{th}=50$~mV, well below $I_cR=61.25$~mV, provided reliable discrimination between the two voltage states. At 100~mK, the switching probability $P_{ sw}$ followed a sharp S-shaped dependence on $I_{ pulse}$ (Fig.~\ref{fig:7}(a)): a change of $\delta I_{pulse}\approx0.1~\mu$A swept $P_{sw}$ from near zero to near unity. At $I_{ pulse}=0.3512$~mA, $P=0.5033$, yielding a Shannon entropy $H(X)\approx1$ bit per switching event (see Methods for the binary mapping and entropy definition)~\cite{shannon1948mathematical}. The time-domain waveforms (Fig.~\ref{fig:7}(b)) confirm that the switching events are unpredictable and independent.

\begin{figure}[htbp]
\includegraphics[width=1\linewidth]{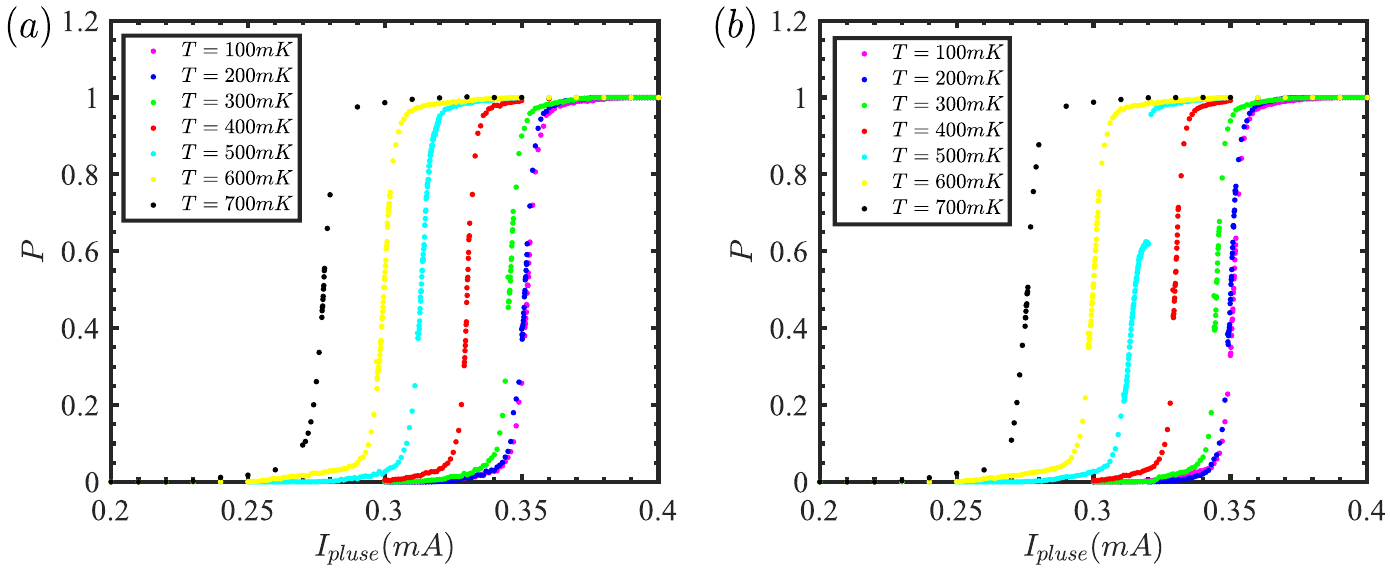}
\caption{\label{fig:8}
JJ switching probability as a function of pulse amplitude at different temperatures. (a) 0.5~kHz; (b) 1~kHz. The temperature range is 100–700~mK, and the pulse duration is $t_p=0.5$~ms.}
\end{figure}

We then examined the temperature dependence from 100~mK to 700~mK (Fig.~\ref{fig:8}), spanning the full range from the base temperature of a dilution refrigerator to the still plate. The S-shaped response remained well-defined at every temperature, and the optimal bias decreased monotonically with temperature (Tab.~\ref{tab:1}), in quantitative agreement with Kramers' theory of thermal escape~\cite{10.1063/1.4824308}. The response is predictable and stable. Across a 600~mK range, over a factor of two in pulse frequency, and despite a three-order-of-magnitude deviation in device parameters, the stochastic switching persisted. This establishes the first key result: the entropy source is the phase-escape process itself, not the nominal device parameters.
\begin{table}[htbp]
\caption{\label{tab:1}
The optimal pulse amplitude obtained under different temperatures and pulse frequencies, and the corresponding switching probability.}
\begin{ruledtabular}
\begin{tabular}{c c c c c}
\multicolumn{1}{c}{Temperature} &
\multicolumn{2}{c}{0.5~kHz} &
\multicolumn{2}{c}{1~kHz} \\
\cmidrule(lr){2-3} \cmidrule(lr){4-5}
& $I_{\rm pulse}$ & $P$ & $I_{\rm pulse}$ & $P$ \\
\hline
$100$~mK & 0.3520 & 0.4910 & 0.3512 & 0.5033 \\
$200$~mK & 0.3512 & 0.4982 & 0.3501 & 0.4895 \\
$300$~mK & 0.3457 & 0.4982 & 0.3450 & 0.5192 \\
$400$~mK & 0.3300 & 0.5050 & 0.3298 & 0.5110 \\
$500$~mK & 0.3133 & 0.4975 & 0.3152 & 0.4980 \\
$600$~mK & 0.2996 & 0.5062 & 0.2998 & 0.5106 \\
$700$~mK & 0.2776 & 0.5017 & 0.2760 & 0.5063 \\
\end{tabular}
\end{ruledtabular}
\end{table}

\subsection{True random number generation}
The preceding results show that even a badly fabricated junction produces robust stochastic switching. The remaining question is whether this switching is random enough to satisfy the NIST SP 800-22 test suite; the international standard for cryptographic randomness. We find that it is, with a margin that renders further device optimization unnecessary.

\begin{figure}[htbp]
\includegraphics[width=0.9\linewidth]{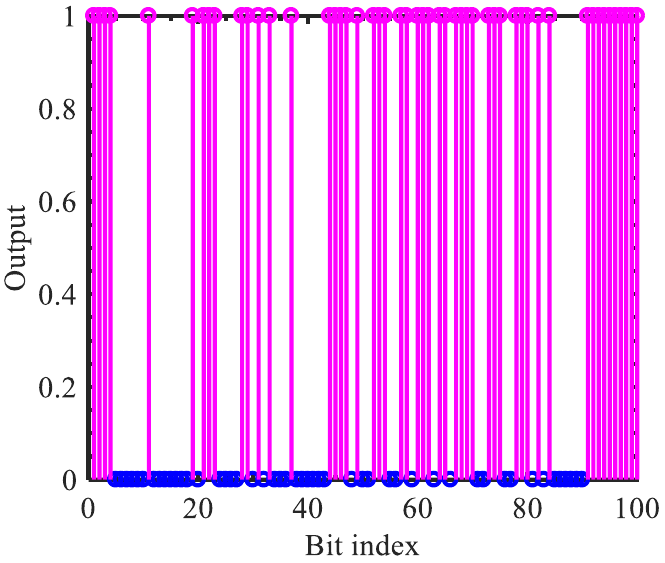}
\caption{\label{fig:9}
Raw random bit sequence measured under conditions of 100~mK, $I_{pulse}=0.3512$~mA, and 1~kHz square wave pulses. Only the first 100 bits are shown in the figure.}
\end{figure}

We acquired two-hour continuous bitstreams from the junction under all 14 conditions listed in Tab.~\ref{tab:1} (seven temperatures from 100~mK to 700~mK, at pulse frequencies of 0.5~kHz and 1~kHz). Fig.~\ref{fig:9} shows a representative raw sequence at 100~mK and 1~kHz. The single-bit statistics were close to ideal: $P(1)=0.5259$, $H(X)=0.9981$~bit, and $H_{min}=0.9271$ bit. At 0.9981 bits, the raw Shannon entropy lies within 0.2\% of the theoretical maximum. The imperfections of the device are, for practical purposes, invisible to the entropy statistics.

The raw sequence did show a first-order autocorrelation $\rho(1)=0.3195$ at 1~kHz; a value that might seem concerning. A closer examination shows that this correlation reflects a clean, well-understood physical process rather than the uncontrolled degradation that plagues other TRNG technologies. Two observations support this interpretation. Lowering the pulse frequency from 1~kHz to 0.5~kHz reduced $\rho(1)$ by a factor of four, from 0.3195 to 0.0764, indicating that the memory is temporal, not structural. The estimated correlation time $\tau_c\approx0.4$~ms is consistent with the junction's RC recovery time, confirming an electrical origin. There is no fatigue, no hysteresis, no state drift, and only a reversible, engineerable RC process. Memristive, spin-torque and magnetic TRNGs suffer from irreversible cycle-to-cycle degradation that accumulates and cannot be calibrated out~\cite{gong2019true}. Our junction, by contrast, displays correlations that disappear when the pulse interval is extended. In this respect, the ``bad'' junction is more stable than the ``engineered'' alternatives.

Tab.~\ref{tab:5} summarizes the raw-bit statistics for all 14 conditions. In 12 of 14 cases, $H(X)>0.97$ bits; in the remaining two, $H(X)>0.86$ bits. Even at the parameter extremes, where optimal biasing becomes more difficult, the entropy remained well above the level that would compromise post-processing. Three systematic patterns emerge from the data. The pulse amplitude can be tuned to $P(1)\approx0.5$ for 12 of 14 conditions; the two outliers occur at the boundaries of the parameter space, not at any fundamental limit. The optimal bias decreases monotonically with temperature, matching Kramers' theory and thus the behaviour is understood and predictable~\cite{10.1063/1.4824308}. And the pulse frequency controls the correlation strength without degrading the entropy: higher frequency gives more bits and more correlation; lower frequency gives fewer bits and less correlation. This is an engineering trade-off, not a physical limitation.
\begin{table*}[htbp]
\caption{\label{tab:5}
Original random sequences and SHA-256 extracted results under different temperatures and driving frequencies. $N$, $P(1)$, $H(X)$, $H_{ min}$, and $\rho(1)$ represent the original sequence length, the probability of ``1", Shannon entropy, min-entropy, and first-order autocorrelation coefficient, respectively; $\tilde{P}(1)$ and $\tilde{\rho}(1)$ represent the corresponding parameters after SHA-256 extraction.
}
\begin{ruledtabular}
\begin{tabular}{ccccccccc}
Temperature & Frequency & $N$ & $P(1)$ & $H(X)$ &
$H_{min}$ & $\rho(1)$ & $\tilde{P}(1)$ & $\tilde{\rho}(1)$\\
\hline
100 mK & 0.5 kHz &3832513&0.4053&0.9740&0.7498&0.0764&0.4998&-0.0001\\
100 mK & 1 kHz &7199857&0.5259&0.9981&0.9271&0.3195&0.4998&-0.0003\\
200 mK & 0.5 kHz &3603706&0.5009&1.0000&0.9974&0.0918&0.5003&-0.0005\\
200 mK & 1 kHz &7199858&0.5026&1.0000&0.9925&0.2701&0.4999&-0.0008\\
300 mK & 0.5 kHz &3609157&0.5023&1.0000&0.9934&0.0470&0.4999&-0.0004\\
300 mK & 1 kHz &7199857&0.5012&1.0000&0.9965&0.1446&0.4998&-0.0003\\
400 mK & 0.5 kHz &3620884&0.5186&0.9990&0.9473&-0.0143&0.4999&0.0011\\
400 mK & 1 kHz &7199858&0.4911&0.9998&0.9745&0.0497&0.4999&0.0002\\
500 mK & 0.5 kHz &3761720&0.4404&0.9897&0.8375&0.0226&0.5005&-0.0013\\
500 mK & 1 kHz &7199856&0.7153&0.8618&0.4834&0.0157&0.5001&-0.0003\\
600 mK & 0.5 kHz &3606478&0.4852&0.9994&0.9579&0.0535&0.5000&0.0001\\
600 mK & 1 kHz &7199860&0.5008&1.0000&0.9977&0.0069&0.4996&0.0000\\
700 mK & 0.5 kHz &3617710&0.4750&0.9982&0.9296&-0.0272&0.5001&0.0003\\
700 mK & 1 kHz &7199856&0.4959&1.0000&0.9882&0.0043&0.5003&0.0004\\
\end{tabular}
\end{ruledtabular}
\end{table*}

To remove the residual bias and short-range correlations, we applied SHA-256 as an entropy extractor, grouping the raw bits into 512-bit blocks and compressing each to a 256-bit output (50\% compression). The hash function does not generate entropy; it concentrates the physical entropy already present. The success of the extraction is therefore direct evidence that the raw entropy is genuine. After extraction, across all 14 conditions, $\tilde{P}(1)$ lay within 0.0005 of 0.5 and $\tilde{\rho}(1)$ fell to $10^{-3}$, a four-order-of-magnitude reduction in correlation.

We then submitted the extracted sequences to the NIST SP 800-22 test suite~\cite{bassham2010sp}. All sequences passed all applicable tests. This result has three notable aspects. First, the pass was not marginal: at 100~mK and 1~kHz, all 15 tests including Runs, Approximate Entropy, and Serial, passed with $P$-values comfortably above $\alpha=0.01$. Second, the pass extended across all operating conditions: sequences from all seven temperatures at 1~kHz passed (Tab.~\ref{tab:6}). Third, the evaluation used a deliberately conservative criterion. With $m=3$ independent sequences of $10^6$ bits, any single failure would have given a ``not pass'' outcome. The standard NIST recommendation of 100 sequences allows a few failures within statistical tolerance; our $m=3$ criterion allows none. The fact that all sequences passed under this stricter standard provides unusually strong evidence of statistical integrity. For the Random Excursions and Random Excursions Variant tests, which require specific preconditions, only one sequence met the conditions; that sequence passed.
\begin{table}[htbp]
\begin{ruledtabular}
\caption{\label{tab:6}
NIST SP 800-22 test results of SHA-256 extracted sequences under 1~kHz driving conditions at different temperatures.}
\begin{tabular}{cccc}
Temperature&Frequency&Bits length&Pass rate\\
\hline
100~mK&1~kHz&3599872&All Pass\\
200~mK&1~kHz&3599872&All Pass\\
300~mK&1~kHz&3599872&All Pass\\
400~mK&1~kHz&3599872&All Pass\\
500~mK&1~kHz&3599872&All Pass\\
600~mK&1~kHz&3599872&All Pass\\
700~mK&1~kHz&3599872&All Pass\\
\end{tabular}
\end{ruledtabular}
\end{table}

These NIST results, together with the near-ideal raw entropy ($H=0.9981$, within 0.2\% of the theoretical maximum), provide a dual validation: the entropy source is both robust in practice and certified by the international standard. A junction that would be rejected by any SFQ designer thus passes the NIST gold-standard test under a criterion more demanding than the standard recommendation. The results demonstrate that substantial fabrication-induced parameter variations can be tolerated without compromising the entropy-generation performance.

Tab.~\ref{tab:7} compares our approach with representative TRNG architectures. The relevant comparison is not output rate but integration compatibility. CMOS and optical TRNGs operate at room temperature and cannot be placed inside a dilution refrigerator. SFQ-based and chaos-based superconducting TRNGs operate at cryogenic temperatures but require extensive peripheral circuitry and, crucially, tightly specified junction parameters~\cite{zhou200150,onomi2020hardware,mizugaki2024operation,oikawa2024chaotic,shimakage2014chaotic}. A device with our parameter deviations would be rejected outright in any SFQ implementation. Our architecture tolerates such deviations and renders parameter matching irrelevant. Where SFQ designers must screen, trim, and match at considerable cost, we require nothing beyond a functional junction with measurable phase-escape dynamics. The robustness and certified randomness are, to our knowledge, unprecedented among cryogenic TRNGs.
\begin{table*}[htbp]
\caption{\label{tab:7}
Characteristic comparison of TRNG schemes with different physical entropy sources.}
\begin{ruledtabular}
\begin{tabular}{ccccc}
Plan Type & Entropy source & Work environment & Output rate & Main System Components \\
\hline
CMOS TRNG~\cite{matsumoto20081200mum}
& Thermal noise
& 300~K
& Mbps-Gbps
& CMOS Circuits \\

Optical TRNG~\cite{herrero2017quantum}
& Photon fluctuations
& 300~K
& Gbps and above
& Light Source, Optical Path, and Detector \\

SFQ~\cite{zhou200150,mizugaki2024operation}
& Superconducting circuits
& Low temperature
& Mbps--Gbps
& Superconducting logic circuits \\

JJ Chaos TRNG~\cite{oikawa2024chaotic,shimakage2014chaotic}
& Chaotic Dynamics
& Low temperature
& High-speed
& JJ and Radio Frequency Excitation System \\

This work
& Phase escape dynamics
& 100-700~mK
& $\sim$1~kbit/s
& Single JJ with pulse bias/readout circuit \\
\end{tabular}
\end{ruledtabular}
\end{table*}
\section{Discussion}

The data lead to a concise chain of inference. A deliberately flawed junction, a device that any conventional standard would reject, can produce near-optimal entropy ($H=0.9981$, NIST-certified). There is no significant gap between this worst-case device and the theoretical upper bound for any binary entropy source ($H=1.0000$). A carefully optimized junction, with tightly controlled parameters, therefore cannot produce measurably better entropy. The entropy quality is already saturated at the level reached by our non-ideal device, and that saturation itself is the demonstration of robustness. If a severely non-ideal device performs at near-optimal levels, then device quality is not a limiting factor for this application. Further tightening of fabrication tolerances buys no improvement in entropy quality.

Would a junction with ideal parameters not be more trustworthy? The NIST test does not have an ``A+'' grade; a sequence either passes or fails. Our sequences pass, with margins well above threshold. There is no higher certification to aim for. The raw Shannon entropy of 0.9981 bits is within 0.2\% of the fundamental maximum of 1 bit; no binary source can do better. And the correlation in the raw bitstream is not a material defect but a reversible, engineerable RC process, and a ``good'' junction with the same capacitance and resistance would also exhibit it. The correlation, like the entropy itself, is essentially independent of device parameters.

The origin of this saturation is clear from Eq.~\eqref{eq:P}. The switching probability can be tuned to $P\approx0.5$ for any junction by adjusting the bias current, because $\Gamma(I_b)$ depends exponentially on the barrier height. The critical current $I_c$ is merely a scaling factor: a tenfold variation in $I_c$ is offset by a proportional change in $I_{pulse}$. The phase-escape mechanism is universal across Josephson junctions, independent of quantitative device parameters. A ``good'' junction shifts the switching curve to a different current scale; it does not make the curve sharper, the entropy higher, or the NIST pass rate better. The entropy source is the phase-escape process, not the device parameters.

The manufacturing implications are substantial. Screening is unnecessary: ``bad'' junctions perform as well as ``good'' ones. Trimming is unnecessary: variations in device parameters do not affect entropy quality. Tight process control is unnecessary: the return on investment in fabrication optimization is zero. This is not an incremental improvement but a fundamental cost reduction. A TRNG can be built from junctions that would otherwise be discarded; the effective fabrication yield is maximized, not reduced.

We note that this study is a proof-of-principle demonstration. The data volume, though sufficient to establish NIST-certified quality and the saturation argument, is constrained by the measurement time available in a dilution refrigerator. A production-grade implementation with the full 100-sequence NIST validation is an engineering task, not a question of physical feasibility. Faster pulsing, optimized readout and circuit-level engineering of the recovery dynamics can raise the bit rate toward the Josephson plasma frequency ($\gtrsim$GHz) while preserving the core advantage: a single-junction footprint, zero static power, and a tolerance to parameter variations that makes screening obsolete.

More broadly, this work suggests a design principle for cryogenic hardware security: the most scalable entropy sources are not those with the most tightly specified components, but those whose randomness arises from physical processes that are inherently tolerant to fabrication imperfections. The Josephson phase-escape dynamics exemplify this principle. The worst junctions generate the best randomness; not because they are special, but because the underlying mechanism is universal and its performance is saturated. Further device optimization yields no measurable return. The practical message is simple: embrace imperfection. It is already good enough, and in this case, good enough is indistinguishable from perfect.
\section{Methods}
\subsection{RCSJ model and phase-escape theory}

The dynamics of a CBJJ are described by the resistively and capacitively shunted junction (RCSJ) model~\cite{devoret1985measurements,clarke1988quantum}:
\begin{equation}
\frac{\hbar C}{2e}\frac{d^2\varphi}{dt^2}
+\frac{\hbar}{2eR}\frac{d\varphi}{dt}
+I_c\sin(\varphi)=I_b,
\end{equation}
where $C$, $R$, and $I_b$ are the junction capacitance, normal-state resistance, and bias current, respectively, and $\varphi$ is the superconducting phase difference. The junction's phase dynamics map onto a particle of mass $m=C(\hbar/2e)^2$ moving in a tilted washboard potential $U(\varphi)=E_J[1-\cos\varphi-(I_b/I_c)\varphi]$, where $E_J=\hbar I_c/2e$ is the Josephson energy. For $I_b<I_c$, the potential has local minima separated by barriers of height $\Delta U(I_b)=2E_J[\sqrt{1-(I_b/I_c)^2}-(I_b/I_c)\arccos(I_b/I_c)]$.

The total escape rate from a potential well is $\Gamma_{ total}=\Gamma_T+\Gamma_Q$, where the thermal escape rate is given by Kramers' theory~\cite{10.1063/1.4824308}:
\begin{equation}
\Gamma_T(I_b)=\frac{\omega_p}{2\pi}a_t\exp\left(-\frac{\Delta U(I_b)}{k_BT}\right),
\end{equation}
with $\omega_p$ the plasma frequency and $a_t$ a prefactor of order unity. The quantum escape rate, dominated by macroscopic quantum tunneling, is~\cite{caldeira1981influence}:
\begin{equation}
\Gamma_Q(I_b)=\frac{\omega_p}{2\pi}a_q\exp\left[-\frac{\Delta U(I_b)}{\hbar\omega_p}\right].
\end{equation}
For the experimental conditions in this work ($T\gtrsim100$~mK and pulse repetition frequencies $\lesssim1$~kHz), thermal activation is the dominant escape mechanism, though quantum tunneling contributes measurably at the lowest temperatures.
\subsection{Device fabrication}
\begin{figure}[htbp]
\includegraphics[width=0.9\linewidth]{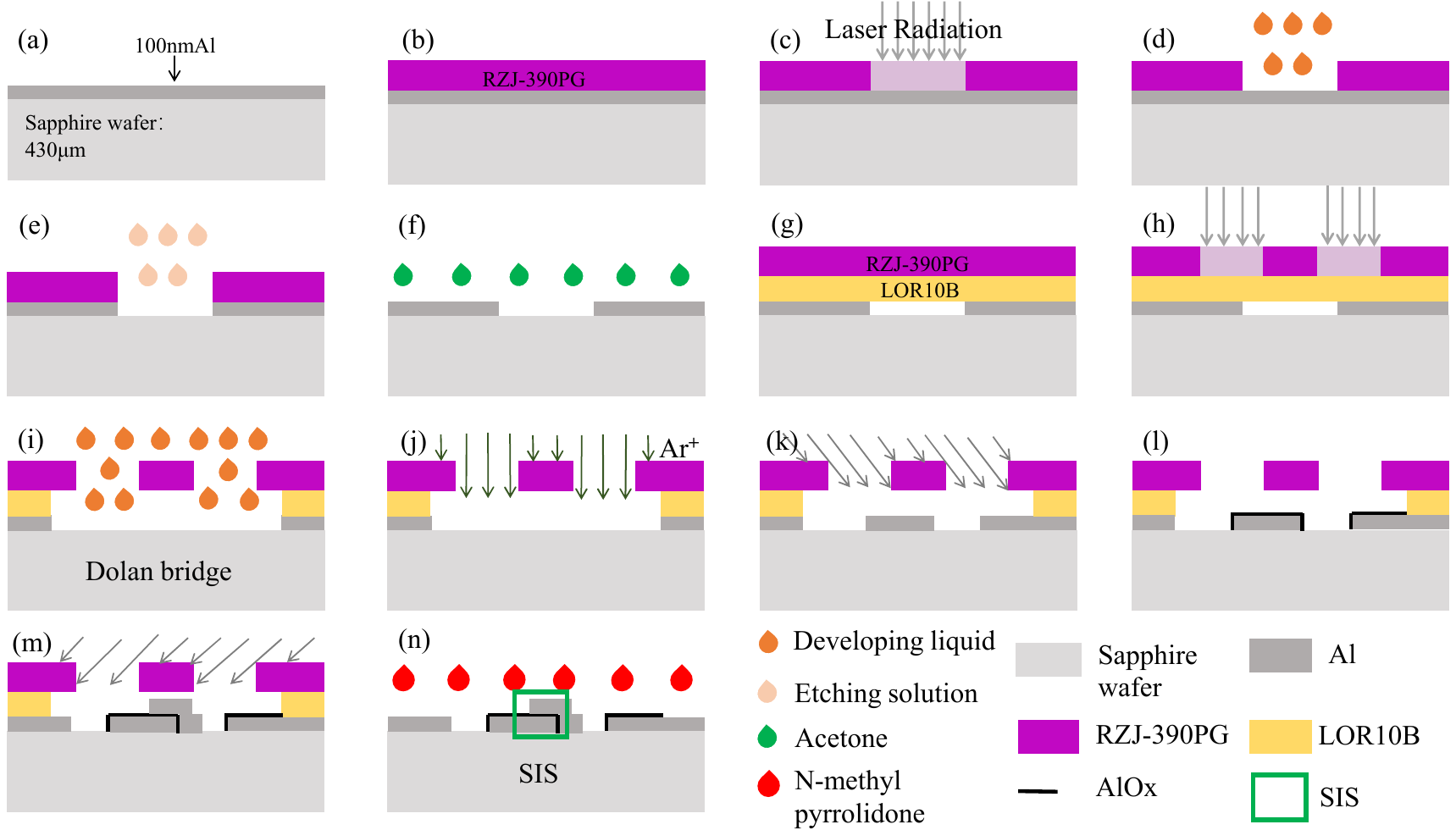}
\caption{\label{fig:JJ-make}
Preparation process of Al/AlOx/Al JJ.
}
\end{figure}

Al/AlOx/Al Josephson junctions were fabricated on sapphire substrates using electron-beam lithography and the Dolan-bridge double-angle evaporation process (Fig.~\ref{fig:JJ-make}). The base pressure during electron-beam evaporation was $\sim10^{-5}$~Pa, with deposition rates of 0.5~nm/s for the bottom electrode and 0.5-0.6~nm/s for the two junction layers. The AlOx tunnel barrier was formed by static oxidation at 1~Pa O$_2$ for 30~min. Transmission-line widths and gaps were approximately 14~$\mu$m, and the junction area was approximately 1~$\rm \mu m^2$.

\subsection{Electrical characterization}

\begin{figure}[htbp]
\includegraphics[width=0.9\linewidth]{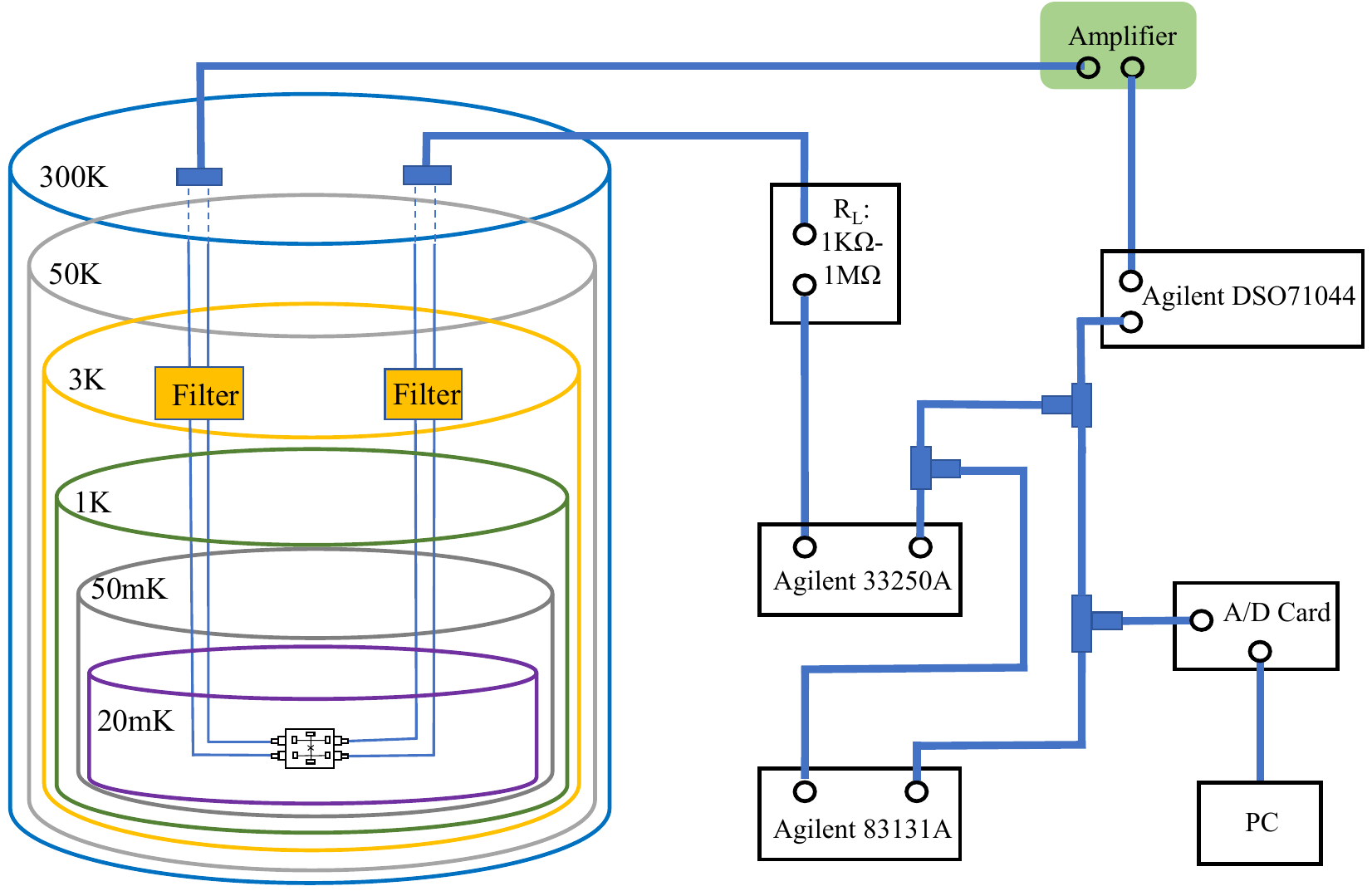}
\caption{\label{fig:JJ_IV}
JJ I-V characteristic measurement circuit. The bias current is applied to the JJ through a high-resistance resistor, and the voltage across the junction is collected after low-temperature filtering and room-temperature amplification.
}
\end{figure}

Electrical characterization was performed in a dilution refrigerator with the sample stage temperature controlled between 100~mK and 700~mK (measurement circuit in Fig.~\ref{fig:JJ_IV}). The bias current was supplied through a large series resistor $R_L\gg R$ (where $R$ is the junction normal-state resistance), ensuring current-bias conditions. Junction voltage was amplified at room temperature and recorded by a digital oscilloscope. The critical current $I_c$ was identified as the current at which the junction transitioned from the zero-voltage to the finite-voltage state, and the normal-state resistance $R$ was extracted from the linear region of the I-V characteristic.
\subsection{Switching probability measurement}

To characterize the stochastic switching behavior, we repeatedly swept the bias current with a triangular waveform at a constant sweep rate $dI_b/dt=0.168$~A/s. Each sweep started from zero bias and increased linearly until a voltage switching event was detected (threshold $V_{th}=50$~mV, set well below $I_cR$), at which point the bias was rapidly reset to zero to allow the junction to relax before the next sweep. The switching current $I_{sw}$ for each event was recorded as $I_{sw}=(dI_b/dt)t_V$, where $t_V$ is the time elapsed until switching. Repeating this cycle 10,000 times yielded the switching-current distribution shown in Fig.~\ref{fig:5}(b).

For probability-tuned operation, we applied square-wave current pulses with duration $t_p=0.5$~ms and a 50\% duty cycle. The pulse amplitude $I_{pulse}$ (always below $I_c$) was adjusted to control the switching probability. For each pulse, the junction either switched to the finite-voltage state (bit ``1'') or remained in the zero-voltage state (bit ``0''), with the outcome determined by comparing the junction voltage against $V_{th}=50$~mV. The switching probability was calculated as $P(1)=N_1/N$, where $N_1$ is the number of ``1'' outputs and $N$ the total number of pulses. We optimized $I_{pulse}$ at each temperature and repetition frequency to minimize $|P(1)-0.5|$ using a two-step optimization: a coarse scan to identify the probability range, followed by a fine scan with 0.1~$\mu$A steps to pinpoint the operating point.
\subsection{Entropy and Correlation Analysis of the Original Random Sequence}
Perform statistical analysis of the experimentally obtained raw bit sequence to evaluate the physical entropy generated by stochastic voltage-state switching of the JJ. First, calculate the Shannon entropy of the raw random bit sequence. For a binary random variable X with a probability $p$ of outputting bit ``1", its Shannon entropy is defined as
\begin{equation}
H(X)
=
-p\log_2p
-(1-p)\log_2(1-p).
\end{equation}
At the same time, the minimum entropy is used to evaluate the worst-case uncertainty of a single sample in the original random sequence, which is defined as
\begin{equation}
H_{\mathrm{min}}(X)=-\log_2\left[\max(P(0),P(1))\right].
\end{equation}
In addition, calculate the first-order autocorrelation coefficient of the original random sequence to evaluate the statistical correlation between adjacent output bits:
\begin{equation}
\rho(1)=\frac{\left\langle(b_i-\mu)(b_{i+1}-\mu)\right\rangle}{\sigma^2},
\end{equation}
Here, $\mu$ and $\sigma$ are the mean and standard deviation of the random bit sequence, respectively. The closer $\rho(1)$ is to zero, the weaker the linear correlation between adjacent bits.
\subsection{Random bit generation and post-processing}

For each operating condition (temperature and pulse frequency), we acquired raw random bit sequences continuously for 2~h. The single-bit probability $P(1)$, Shannon entropy, min-entropy, and first-order autocorrelation coefficient were computed for the raw sequences (following standard information-theoretic definitions). To remove statistical biases and short-range correlations, we applied SHA-256 cryptographic hashing as an entropy extractor: consecutive raw bits were grouped into 512-bit blocks, and each block was hashed to a 256-bit output, yielding a 50\% compression ratio. The hashing process extracts existing physical entropy and does not introduce entropy of its own.

\subsection{Randomness evaluation}

NIST SP 800-22 statistical tests were performed on the SHA-256-extracted sequences at a significance level of $\alpha=0.01$~\cite{bassham2010sp}. In this proof-of-principle demonstration, we divided the acquired data into $m=3$ independent sequences of $10^6$ bits each—a sample size that, while smaller than the 100-sequence threshold recommended for a full $P$-value uniformity assessment, provides a stringent and conservative initial validation of the entropy source's statistical quality. With only three sequences, any single failure among them would have constituted a test failure. The fact that all sequences passed under this limited dataset provides strong evidence that the observed randomness is genuine and not an artifact of statistical leniency. For tests with additional preconditions (Random Excursions, Random Excursions Variant), we report results only for sequences meeting the NIST-specified conditions. We note that while this sample size is insufficient for the standard uniformity test on $P$-values, it is sufficient to provide a rigorous lower-bound validation of entropy quality.
\subsection{Prototype demonstration of the JJ-TRNG}
\begin{figure}[htbp]
\includegraphics[width=1\linewidth]{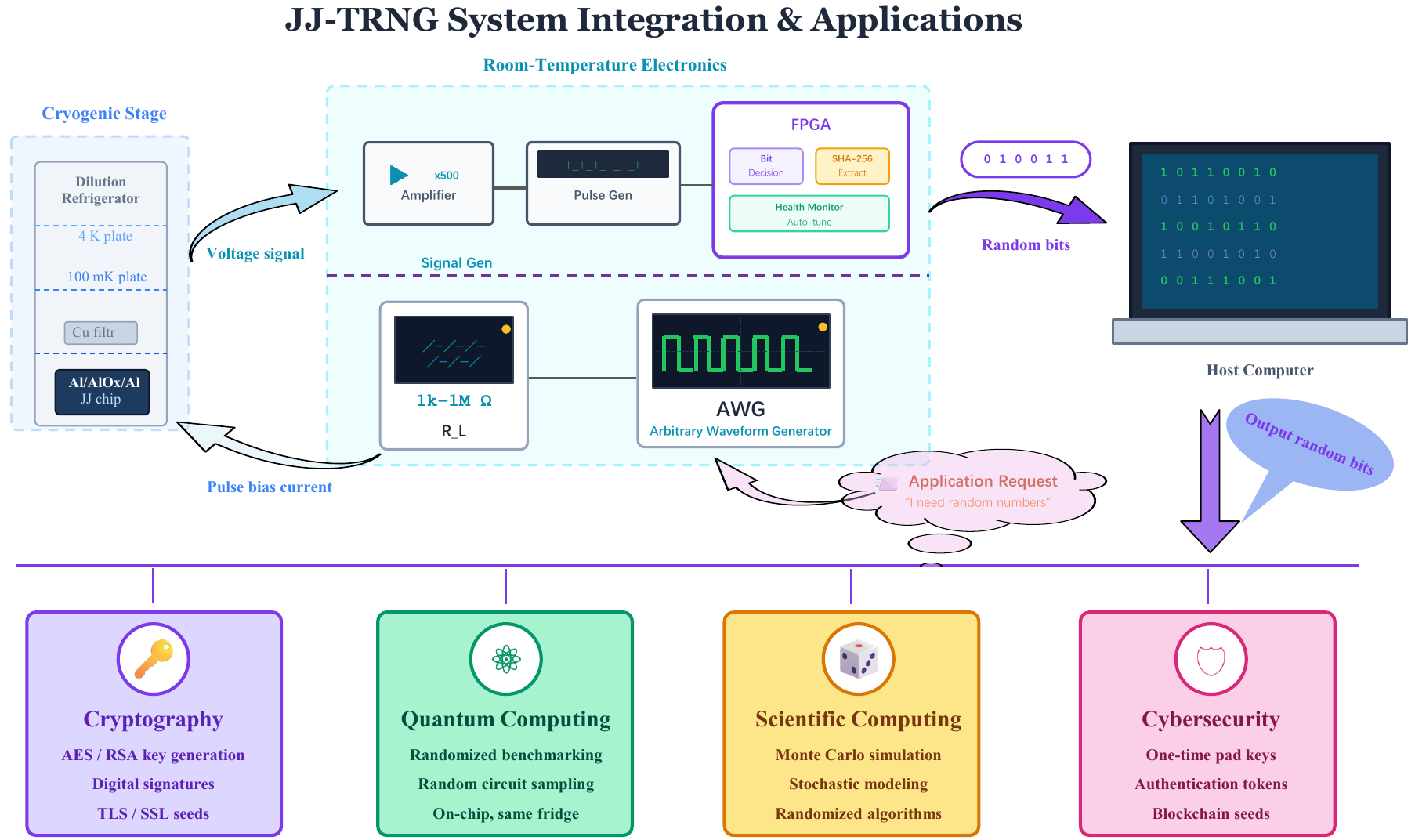}
\caption{\label{fig:10}
Schematic diagram of a true random number generation scheme based on a single Josephson junction. Random voltage state switching of the JJ is driven by pulse bias, and random bits are obtained through voltage decision and SHA-256 entropy extraction.
}
\end{figure}

To demonstrate the practical operation of the JJ-TRNG, we implement a single-junction prototype based on stochastic voltage-state switching, as illustrated in Fig.~\ref{fig:10}. At the selected operating temperature, the pulse amplitude is adjusted to bring the switching probability close to $P=0.5$. Periodic current pulses are then applied to induce stochastic switching of the junction between the voltage and zero-voltage states. These states are digitized into raw binary bits and subsequently processed using SHA-256-based entropy extraction before statistical evaluation.
\begin{acknowledgments}
L. F. Wei discloses support for the research of this work from the National Key Research and Development Program of China [grant number 2021YFA0718803]. X. N. Feng discloses support for the research of this work from the National Natural Science Foundation of China [grant number 11974290].
\end{acknowledgments}

\bibliography{cite}

\end{document}